\documentclass[twocolumn,english,pra,amsmath,amssymb,superscriptaddress,showpacs]{revtex4}
\usepackage[T1]{fontenc}
\usepackage[latin9]{inputenc}
\usepackage{babel}

\usepackage{amsmath}
\usepackage{graphicx}
\usepackage{amssymb}
\usepackage{esint}
\usepackage[unicode=true, pdfusetitle,
 bookmarks=true,bookmarksnumbered=false,bookmarksopen=false,
 breaklinks=false,pdfborder={0 0 1},backref=false,colorlinks=false]
 {hyperref}

\makeatletter

\newcommand{\lyxdot}{.}

\@ifundefined{textcolor}{}
{%
 \definecolor{BLACK}{gray}{0}
 \definecolor{WHITE}{gray}{1}
 \definecolor{RED}{rgb}{1,0,0}
 \definecolor{GREEN}{rgb}{0,1,0}
 \definecolor{BLUE}{rgb}{0,0,1}
 \definecolor{CYAN}{cmyk}{1,0,0,0}
 \definecolor{MAGENTA}{cmyk}{0,1,0,0}
 \definecolor{YELLOW}{cmyk}{0,0,1,0}
 }

\usepackage{mathrsfs}

\makeatother

\makeatother

\makeatother

\begin{document}

\title{Quantum correlations in a cluster-like system}

\author{Yi-Xin Chen}

\email{yxchen@zimp.zju.edu.cn}

\affiliation{Zhejiang Institute of Modern Physics, Zhejiang University, Hangzhou
310027, China}

\author{Sheng-Wen Li}

\email{swli@zimp.zju.edu.cn}

\affiliation{Zhejiang Institute of Modern Physics, Zhejiang University, Hangzhou
310027, China}

\author{Zhi Yin}

\email{zhiyin@zimp.zju.edu.cn}

\affiliation{Zhejiang Institute of Modern Physics, Zhejiang University, Hangzhou
310027, China}

\affiliation{College of Science, Ningbo University of Technology, Ningbo 315000,
China}
\begin{abstract}
We discuss a cluster-like 1D system with triplet interaction. We study
the topological properties of this system. We find that the degeneracy
depends on the topology of the system, and well protected against
external local perturbations. All these facts show that the system
is topologically ordered. We also find a string order parameter to
characterize the quantum phase transition. Besides, we investigate
two-site correlations including entanglement, quantum discord and
mutual information. We study the different divergency behaviour of
the correlations. The quantum correlation decays exponentially in
both topological and magnetic phases, and diverges in reversed power
law at the critical point. And we find that in TQPT systems, the global
difference of topology induced by dimension can be reflected in local
quantum correlations.
\end{abstract}


\maketitle

\section{Introduction}

Nowadays, quantum correlation has been attracting much attention since
it plays a crucial role in quantum computation and quantum information.
Entanglement, as an important quantum resource, takes responsibility
for most quantum information tasks such as quantum teleportation and
computation \cite{horodecki_quantum_2009}. But entanglement is fragile
in open systems. Environment induced decoherence destroys entanglement
correlation in a short time, which makes quantum task difficult for
implementation. 

However, recent reseach shows that entanglement may be not the only
worker carrying on quantum tasks. The quantum correlation without
entanglement may also take effect in some scenes, e.g., the quantum
computation with mixed states plus one pure qubit (DQC1) \cite{Knill,datta_quantum_2008,Fanchini2010}.

Quantum discord is developed for the measure of {}``quantumness''
of a pairwise correlation \cite{ollivier_quantum_2001}. It makes
it clear that entanglement is one kind of nonclassical correlation
but not all. The quantum discord of some separable states is also
nonzero. It may be used as a powerful tool to study quantum correlations.

Lots of work has been devoted to the study of correlations in different
processes, like decoherence and quantum phase transition \cite{dillenschneider_quantum_2008,fanchini_non-markovian_2009,sarandy_classical_2009,chen_quantum_2010,thermal_discord_2010,XXZ_discord_2010,geometric_discord_2010,discord_MPS_2010}.
The entanglement of formation \cite{wootters_1998} does not behave
smoothly like the correlation functions, and shows sudden death and
rebirth in some scences \cite{SuddenBirth_2008}, which attracts more
and more researchers. The quantum discord is pointed out to signal
the quantum phase transition \cite{sarandy_classical_2009} like fidelity
\cite{ma_many-body_2009}, while our previous work also find that
in topological quantum phase transition (TQPT) the local correlations
are classical and the quantum correlation hides in the global system
\cite{chen_quantum_2010}.

The topological order is a new kind of order beyond the conventional
symmetry-breaking theory. In topological order system, the degeneracy
of the ground space depends on the topology of the system configuration,
and the degenerate ground space is well protected against local perturbation.
Such properties can be used for fault-tolerant computation \cite{kitaev_fault-tolerant_2003,nayak_non-abelian_2008,chen_JJATQC_2010}.
Another talent idea is the measurement-based computation, in which
a \emph{cluster state} is prepared and measured as the computation
process \cite{briegel_persistent_2001}. There are deep relationship
between these two methods of computation.

In this paper, we study the pairwise correlations in a 1D cluster-like
system with triplet interaction, which can be implemented in optical
lattice \cite{kay_quantum_2004}. We discuss the properties of the
topological order in the system, like the boundary dependent degeneracy
and topological protection. We find the \emph{string order parameter}
(SOP) by the method of duality map \cite{Fradkin1978,feng_topological_2007}
to characterize TQPT.

Furthermore, the system can be decomposed as two independent chains
of odd and even sites respectively, namely, the spin on site $i$
is independent of spin on site $i+(2n+1)$ where $n$ is integer,
and we call this {}``bridge correlated''. 

The divergency of quantum discord with the distance of two site is
studied. We find that it behaves in the similar way as the correlation
functions, i.e., it decays exponentially in both topological and magnetic
phase areas and diverges in reversed power law at the critical points.

Moreover, the study of the quantum discord and entanglement shows
that the local quantum correlation of two sites is suppressed in topological
phase area. This is different from the study in 2D TQPT \cite{chen_quantum_2010},
in which local quantum correlations vanish completely. And that means
in TQPT systems the global difference of the topology caused by dimension
can be reflected in the local quantum correlations. 

The paper is organized as follows. In Sec. II, we show the basic model
of this cluster-like system. we discuss the topological properties
like the degeneracy of the ground space and the topological protection.
In Sec. III, we study the quantum correlation of this model. We investigate
quantm discord, correlation length and mutual information. Finally,
we draw summary in Sec. V.

\section{Topological properties of cluster-like system}

In this section, we introduce the 1D cluster-like system originally
proposed for quantum computation in optical lattice \cite{kay_quantum_2004}.
We calculate the basic property of low-energy spectrum. In 1D world,
there are not too many kinds of different topology that we are interested
in, except for the open string and the closed loop, which correpond
to open and periodic boundary conditions respectively. We show that
the degeneracy of the ground space is different in these two cases.
Besides, the degeneracy is immune to local perturbation. These are
the typical characters of topological order. And we find the SOP to
characterize the quantum phase transition.

\subsection{The model of cluster-like system}

Here, we describe the model we discussed. The Hamiltonian of the system
is described as follows, \begin{equation}
H=-J\sum_{i}(\sigma_{i-1}^{x}\sigma_{i}^{z}\sigma_{i+1}^{x}+B\sigma_{i}^{z})\equiv-J\sum_{i}(S_{i}+B\sigma_{i}^{z}),\label{eq:H}\end{equation}
 where $J>0$, $\sigma_{i}^{\alpha}$ is the Pauli matrix acting on
the $i$-th site and $S_{i}=\sigma_{i-1}^{x}\sigma_{i}^{z}\sigma_{i+1}^{x}$. 

\begin{figure}
\includegraphics[width=6cm]{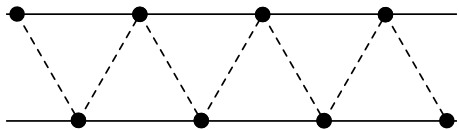}

\caption{Configuration of the system in optical lattice implementation. Tunneling
between the nearest three sites (black points) in a triangle gives
rise to the triplet interaction term.}

\label{fig:Tlattice} 
\end{figure}

The model is originally proposed in Ref. \cite{kay_quantum_2004}
for quantum computation. It can be implemented in optical lattice
. Atoms are arranged in a triangle lattice as shown in Fig. \ref{fig:Tlattice}.
Tunneling happens in the nearest three sites, which gives rise to
the triplet interaction term. The one body term can be adjusted by
Zeeman effect and appropriate laser field.

When $B=0$, the ground space is the common eigenspace of $S_{i}$'s
that satisfies $S_{i}|\mbox{g.s.}\rangle=|\mbox{g.s.}\rangle$, which
is a typically cluster-like space \cite{son_entanglement_2010}. Cluster
states are a kind of graph states, which play an essential role in
measurement-based quantum computation \cite{briegel_persistent_2001}.

Here we analyze the low energy spectrum of the system Eq. (\ref{eq:H})
in the stablizer scheme \cite{nielsen_quantum_2000,chen_JJATQC_2010}.
All $S_{i}$'s commute with each other, so we can treat the ground
space of the system as the protected space of a set of independent
\emph{stabilizer genarators} $\left\{ S_{i}\right\} $.

In the stabilizer scheme, we have $N$ qubits and $K$ independent
stablizer genarators which are the products of Pauli operator $\sigma_{i}^{\alpha}$'s.
The stablizer genarators commute with each other, and the common eigenspace
of the stablizers satisfying $S_{i}|\Phi\rangle=|\Phi\rangle$ is
called the \emph{protected space}, whose dimension is just $2^{N-K}$,
i.e., the stabilizers encode $N-K$ logical qubits in the protected
space.

Assume we have $N$ sites in the system Eq. (\ref{eq:H}), when $B=0$,
it is easy to see that the number of $S_{i}$'s is $N$ under periodic
boundary condition, and $N-2$ under the open boundary condition.
the dimensions of the protected space of $\left\{ S_{i}\right\} $
are $2^{0}=1$ and $2^{2}=4$ respectively. That is to say, the ground
space of the system Eq. (\ref{eq:H}) is undegenerated under the loop
boundary condtion, and 4-fold degenerated when it is opened, with
the absent of external field term.

Another important property of the system is that all the local correlation
functions, except those composed of products of several $S_{i}$'s,
are zero. As an example, for $\hat{o}=\sigma_{i}^{\alpha}\sigma_{j}^{\beta}$,
we can always find certain $S_{k}$ which anti-commutes with $\hat{o}$,
so that $\left\langle \hat{o}\right\rangle =\left\langle \hat{o}S_{k}\right\rangle =-\left\langle S_{k}\hat{o}\right\rangle =-\left\langle \hat{o}\right\rangle =0$.
And the system possesses $Z_{2}$-symmetry, i.e., $\left[H,\prod\sigma_{i}^{z}\right]=0$.
These are important properties as we will see below.

When $B\neq0$, by means of Jordan-Wigner transformation \begin{equation}
\begin{cases}
\sigma_{i}^{x} & =(c_{i}^{\dagger}+c_{i})\prod_{j<i}(1-2c_{j}^{\dagger}c_{j}),\\
\sigma_{i}^{z} & =2c_{i}^{\dagger}c_{i}-1,\end{cases}\label{eq:JW}\end{equation}
 we can transform the system Eq. (\ref{eq:H}) into a fermion chain,
\begin{equation}
-H/J=\sum_{i}(c_{i-1}-c_{i-1}^{\dagger})(c_{i+1}^{\dagger}+c_{i+1})+B(2c_{i}^{\dagger}c_{i}-1).\label{eq:fermion}\end{equation}

We can see that the system can be regarded as two independent chains
containing odd and even sites respectively.

Under the periodic boundary condition, the system possesses translational
invariance. So it can be diagonalized in Fourier representation,

\begin{equation}
-H/J=\sum_{k}e^{-2ik}(a_{k}-a_{-k}^{\dagger})(a_{-k}+a_{k}^{\dagger})+B(2a_{k}^{\dagger}a_{k}-1),\end{equation}
 where $c_{n}=\sum_{k}e^{ikn}a_{k}/\sqrt{N}$. By using Bogoliubov
transformation, we get the diagonalized Hamiltonian, \begin{equation}
H/J=\sum_{k}\epsilon_{k}(2\gamma_{k}^{\dagger}\gamma_{k}-1),\end{equation}
 where $\epsilon_{k}=(1+B^{2}-2B\cos2k)^{\frac{1}{2}}$, $a_{k}=\cos\theta_{k}\gamma_{k}+i\sin\theta_{k}\gamma_{-k}^{\dagger}$
and $\tan2\theta_{k}=\sin2k/(B-\cos2k)$.

When the string is opened, it is difficult to get the low energy spectrum
and we will discuss the degeneracy of the ground space by perturbation
method in the following part.

\subsection{Topologically protected degeneracy}

When the string is opened, the ground space is 4-fold degenerated
when $B=0$, as mentioned above. Actually each independent chain contributes
two states. In this part, we show that this degeneracy is protected
against external local perturbations. More exactly speaking, the energy
splitting caused by perturbation tends to zero in the thermodynamical
limit.

As the string is opened, Fourier transformation does not take effect.
We calculate the splitting of the ground state energy. Assume the
external field is absent at the time $t\rightarrow-\infty$, and adiabatically
switched on. That is to say, we construct a new Hamiltonian with time-dependent
external field $\lambda(t)=e^{-\left|\eta\right|t}$, where $\eta$
is infinitely small. At $t=0$, the system comes back to Eq. (\ref{eq:H}).
That is, \begin{equation}
H(t)=H_{0}+\lambda(t)H'.\end{equation}
 Since $\lambda(t)$ is switched on adiabatically, the system evolves
from the cluster-like ground state $|\Phi_{0}\rangle$ at $t\rightarrow-\infty$
to an eigenstate $|\mathsf{G}\rangle$ of Eq. (\ref{eq:H}) at $t=0$,
which should be one of the splitted ground states \cite{Gell-Mann1951,brouder_many-body_2009}.
The average energy of the state $|\mathsf{G}\rangle$ is,\begin{align}
 & \left\langle \mathsf{G}|H(t=0)|\mathsf{G}\right\rangle \nonumber \\
 & =\left\langle \Phi_{0}|U^{\dagger}(0,-\infty)(H_{0}+H')U(0,-\infty)|\Phi_{0}\right\rangle .\label{eq:energy}\end{align}
 $U(0,-\infty)=\mathsf{T}\exp\left[-i\int_{-\infty}^{0}H'^{I}(t)dt\right]$
is the time-ordered evolution operator, expanded as

\begin{align}
U(0,-\infty) & =\mathbf{1}+(-i)\int_{-\infty}^{0}dtH'^{I}(t)\\
 & +(-i)^{2}\int_{-\infty}^{0}dt_{1}\int_{-\infty}^{t_{1}}dt_{2}H'^{I}(t_{1})H'^{I}(t_{2})+\dots,\nonumber \end{align}
 where the perturbation term in interaction picture is \begin{align}
H'^{I}(t) & =\lambda(t)e^{iH_{0}t}H'e^{-iH_{0}t}\nonumber \\
 & =\lambda(t)\sum e^{-iJt(S_{i-1}+S_{i+1})}\sigma_{i}^{z}e^{iJt(S_{i-1}+S_{i+1})},\end{align}
 ignoring the boundary terms without loss of generality. As $e^{-iJtS_{i}}=\cos Jt-i\sin JtS_{i}$,
we can see that the inner product in Eq. (\ref{eq:energy}) is composed
of sum of multi-point correlation functions, which all vanish until
the $N$th order according to what we see in the last part. In the
$N$th term, global terms like $\left\langle \prod\sigma_{i}^{z}\right\rangle $
appear and take effect. We can interpret it as a virtual particle
running along the whole string. Therefore, the energy splitting of
the ground space is $\sim\exp(-1/L)$, where $L$ is the length scale
of the system. In thermodynamical limit, $L\rightarrow\infty$, the
degeneracy is perfectly protected like the case in toric code \cite{kitaev_fault-tolerant_2003}.

As we see, degeneracy emerges when the loop is opened. Besides, the
degeneracy is immune against local perturbation when it is not too
strong. These properties show that the system is a topologically ordered
system. We can see that there is a close relationship between cluster-like
system and topological order. Here we regard both the topology related
degeneracy and topological protection as the essential character of
topological order.

\subsection{String order parameter}

Topological order is an unconventional phase that cannot be described
by the symmetry-breaking of local order parameters \cite{wen_quantum_2004}.
When $\left|B\right|\rightarrow\infty$, the system leaves the topological
order and goes to a magnetized phase through quantum phase transition.
We can find some global string order parameter to characterize the
phase transition. Bellow, we show how to find the SOP by duality transformation
\cite{Fradkin1978,feng_topological_2007}.

Under the periodic boundary condition, we make such duality transformation
bellow to represent the system by another self-consistent Pauli algebra
$\left\{ \mu_{i}^{\alpha}\right\} $,\begin{equation}
\begin{cases}
\sigma_{i}^{z} & =\mu_{i}^{x}\mu_{i+1}^{x},\\
\sigma_{i}^{x} & =\prod_{j\leq i}\mu_{j}^{z}.\end{cases}\end{equation}
 The system turns to be an $XY$-model, \begin{equation}
-H/J=\sum_{i}-\mu_{i}^{y}\mu_{i+1}^{y}+B\mu_{i}^{x}\mu_{i+1}^{x}.\label{eq:XY}\end{equation}

Further, let $\mu_{i}^{x}=\tau_{i}^{x}\tau_{i+1}^{x}$ and $\mu_{i}^{y}=\prod_{j\leq i}\tau_{j}^{z}$,
the system can be mapped to Ising model in an external field,\begin{equation}
-H/J=\sum_{i}-\tau_{i+1}^{z}+B\tau_{i}^{x}\tau_{i+2}^{x}.\end{equation}
 We can also see that the system is actually composed of two independent
chains. Combining the two transformation together, we can see actually
it is \begin{eqnarray}
\sigma_{i}^{z} & = & \tau_{i}^{x}\tau_{i+2}^{x},\nonumber \\
\sigma_{i-1}^{x}\sigma_{i}^{z}\sigma_{i+1}^{x} & = & \tau_{i+1}^{z}.\end{eqnarray}
 The three nearest sites in a triangle (see Fig. \ref{fig:Tlattice})
make up a new site in the dual lattice. The regular triangles and
the inverted ones construct two independent Ising chains respectively.
$\tau_{i}^{z}$ can be seen as the observable that measure the {}``vortex''
of the $i$th triangle site, clockwised or counter-clockwised.

Lots of work has been devoted to discussing the quantum phase transition
of Ising model. There is a long-range order in the dual system \cite{Pfeuty1970}.
When $\left|B\right|\geq1$, we have\begin{equation}
\lim_{j\rightarrow\infty}\left\langle \tau_{0}^{x}\tau_{2j}^{x}\right\rangle =\left\langle \tau_{2j}^{x}\right\rangle ^{2}\sim\left[1-1/B^{2}\right]^{\frac{1}{4}},\end{equation}
 while vanishes when $\left|B\right|<1$. $\tau_{2j}^{x}$ can be
regarded as the order parameter characterizing the phase transition
at $B=\pm1$. In the original spin representation, we can get the
hidden SOP as \begin{equation}
\Delta_{\mathsf{even(odd)}}=\prod_{i}\sigma_{2i(+1)}^{z}.\end{equation}
 Note that we are treating two independent Ising chains.

When the loop is opened, some boundary terms appears, whose effect
can be neglected in the thermodynamical limit. The physics does not
change.

Here we emphasize that the existence of SOP is not the sufficient
condition of topological order. As we see, we can also get SOP in
the $XY$-model Eq. (\ref{eq:XY}), i.e., $\tau_{0}^{x}\tau_{2j}^{x}=\prod_{i=0}^{2j-1}\mu_{i}^{x}$,
by the duality map, which is a convensional symmetry-breaking system
studied so much. However, duality mapping is a useful tool to help
us find the nonlocal order in topological order system.

\section{pairwise correlations}

In this section, we study the pairwise correlations in our cluster-like
system, like the quantum discord and the entanglement of formation
(EoF). Quantum discord is used as a measure for the {}``quantumness''
of a pairwise state. Something interesting are found. We find that
the quantum correlations are greatly suppressed in the topological
order area compared with the magnetic polarized phase. The quantum
discord decays exponentially as the distance of the two spins increases
when $\left|B\right|\neq1$, and diverges in reverse power law at
critical points, in the behaviour exactly like the two-point correlation
function. Only the EoF of the spins next-nearest is nontrivial, while
that of spins further from each other vanishes.

\subsection{Entanglement, mutual information and quantum discord}

Entanglement, as the most important quantum resource, has been discussed
lots and there are many different kinds of measures. One of the most
sophisticated is the entanglement of formation (EoF) \cite{eof_1996}.
Entanglement of formation is an entanglement measure defined for bipartite
quantum states as

\begin{equation}
E(\rho)\equiv\underset{\{p_{i},|\psi_{i}\rangle\}}{\textrm{min}}\left[\sum_{i}p_{i}S^{E}(|\psi_{i}\rangle)\right],\end{equation}
 where $\rho$ is the density matrix of the bipartite states and $\{p_{i},|\psi_{i}\rangle\}$
satisfies the condition that $\rho=\sum_{i}p_{i}|\psi_{i}\rangle\langle\psi_{i}|$.
$|\psi_{i}\rangle$ is a bipartite pure state and $S^{E}(\cdot)$
gives the von Neumann entropy of the reduced density matrix of $|\psi_{i}\rangle\langle\psi_{i}|$.
For pure states case, this quantity reduces to the entropy of entanglement.
For two-qubit system, fortunately, EoF can be express with concurrence
$C$ \cite{wootters_1998}, \begin{equation}
E(\rho)=-f(C)\log f(C)-(1-f(C))\log(1-f(C)),\end{equation}
 where $f(C)=(1+\sqrt{1-C^{2}})/2$. The concurrence $C=\max[0,\lambda_{1}-\lambda_{2}-\lambda_{3}-\lambda_{4}]$
and $\lambda_{i}$ are the square roots of the eigenvalues of the
matrix $\rho(\sigma^{y}\otimes\sigma^{y})\rho^{*}(\sigma^{y}\otimes\sigma^{y})$.

Mutual information \cite{nielsen_quantum_2000} quanlifies the amount
of common information shared by two subsystems. The classical mutual
information is \begin{eqnarray*}
I(A:B) & = & H(A)+H(B)-H(AB)\\
 & = & H(A)-H(A|B),\end{eqnarray*}
 where $H(\cdot)$ is the Shannon information and $H(A|B)$ is the
conditional information, which means the average information of $A$
we gain when knowing the result of $B$. A natural generalized quantum
version is by change the Shannon information to von Neumann entropy,\begin{equation}
\mathcal{I}(\rho^{AB})=S(\rho^{A})+S(\rho^{B})-S(\rho^{AB}).\label{eq:mi}\end{equation}

Another generalization follows by giving the quantum measurement version
of conditional entropy. The conditional entropy implies a measurement
on $B$ to get the information about $A$. So we impose projective
measurement $\{\hat{\Pi}_{i}^{B}\}$ on $B$ and collect the information,\begin{equation}
\mathcal{J}(\rho^{AB}:\{\hat{\Pi}_{i}^{B}\})=S(\rho^{A})-\sum_{i}p_{i}S(\hat{\Pi}_{i}^{B}\rho^{AB}\hat{\Pi}_{i}^{B}/p_{i}),\end{equation}
 where $p_{i}=\mathrm{Tr}\left[\hat{\Pi}_{i}^{B}\rho^{AB}\hat{\Pi}_{i}^{B}\right]$.

Quantum discord is defined as the minima of the difference of $\mathcal{I}$
and $\mathcal{J}$ \cite{ollivier_quantum_2001}, \begin{equation}
D(\rho^{AB})=\textrm{min}\left[\mathcal{I}(\rho^{AB})-\mathcal{J}(\rho^{AB}:\{\hat{\Pi}_{i}^{B}\})\right].\end{equation}
 Due to its power in mixed state quantum computation \cite{Knill,datta_quantum_2008,Fanchini2010},
it has been discussed a lot recently. 

Quantum discord can be used a measurement for the {}``quantumness''
of the bipartite correlation. It clears that entanglement is not the
only {}``quantum'' state. For example, for a separable state $\rho=|00\rangle\langle00|/2+|++\rangle\langle++|/2$,
where $|+\rangle=(|0\rangle+|1\rangle)/\sqrt{2}$, the quantum discord
is not zero, which means $\rho$ contains nonclassical correlation.

\subsection{Pairwise state}

To study the correlations in the system, we should first get the state
of two spins, i.e., their reduced density matrix. The pairwise density
matrix can be decomposed by a set of basis $\{\frac{1}{2}\sigma_{i}^{\mu}\sigma_{j}^{\nu}\}$,
where $\mu$ and $\nu$ takes $0,\ldots,3$ and $\sigma_{i}^{0}=\mathbf{1}$.
It can be easily checked that $\{\frac{1}{2}\sigma_{i}^{\mu}\sigma_{j}^{\nu}\}$
is orthonormal under the Hilbert-Schmidt inner product $(A,B)_{\mathrm{H-S}}\equiv\mathrm{\mathrm{tr}}(A^{\dagger}B)$
\cite{nielsen_quantum_2000,wang_pairwise_2002,chen_quantum_2010}.
The reduced density matrix of two spins can be written as\begin{equation}
\rho_{ij}=\frac{1}{4}\sum_{\mu\nu}\left\langle \sigma_{i}^{\mu}\sigma_{j}^{\nu}\right\rangle \sigma_{i}^{\mu}\sigma_{j}^{\nu},\label{eq:RDM}\end{equation}
 where $\left\langle \sigma_{i}^{\mu}\sigma_{j}^{\nu}\right\rangle =\mathrm{Tr}(\rho_{\mathsf{G}}\sigma_{i}^{\mu}\sigma_{j}^{\nu})=\mathrm{tr}(\rho_{ij}\sigma_{i}^{\mu}\sigma_{j}^{\nu})$
is the Hilbert-Schmidt inner product of $\rho_{ij}$ and $\sigma_{i}^{\mu}\sigma_{j}^{\nu}$.

Because of the $Z_{2}$-symmetry of the system mentioned before, most
terms in Eq. (\ref{eq:RDM}) can be elimited except that of $\mathbf{1}_{ij}$,
$\sigma_{i}^{\mu}\sigma_{j}^{\mu}$, $\sigma_{i}^{z}\otimes\mathbf{1}_{j}$
and $\mathbf{1}_{i}\otimes\sigma_{j}^{z}$. So we just need to calculate
the expectation value of $\left\langle \sigma_{i}^{\mu}\sigma_{i+R}^{\mu}\right\rangle $
and $\left\langle \sigma^{z}\right\rangle $. Since the system can
be treated as two independent fermion chain like Eq. (\ref{eq:fermion}),
it can be seen that $\left\langle \sigma_{i}^{\mu}\sigma_{i+R}^{\mu}\right\rangle $
is zero when $R$ is odd.

From the Jordan-Wigner transformation Eq. (\ref{eq:JW}), we can get
$\left\langle \sigma^{z}\right\rangle $ and $\left\langle \sigma_{0}^{\mu}\sigma_{R}^{\mu}\right\rangle $
directly (we take $i=0$ without loss of generality).\begin{eqnarray*}
\left\langle \sigma^{z}\right\rangle  & = & \left\langle (c_{0}-c_{0}^{\dagger})(c_{0}+c_{0}^{\dagger})\right\rangle \\
 & = & \left\langle A_{0}B_{0}\right\rangle ,\\
\left\langle \sigma_{0}^{z}\sigma_{R}^{z}\right\rangle  & = & \left\langle (c_{0}-c_{0}^{\dagger})(c_{0}+c_{0}^{\dagger})(c_{R}-c_{R}^{\dagger})(c_{R}+c_{R}^{\dagger})\right\rangle \\
 & = & \left\langle A_{0}B_{0}A_{R}B_{R}\right\rangle ,\\
\left\langle \sigma_{0}^{x}\sigma_{R}^{x}\right\rangle  & = & \langle(c_{0}-c_{0}^{\dagger})(c_{1}+c_{1}^{\dagger})(c_{1}-c_{1}^{\dagger})(c_{2}+c_{2}^{\dagger})\\
 &  & \ldots(c_{R-1}-c_{R-1}^{\dagger})(c_{R}+c_{R}^{\dagger})\rangle\\
 & = & \left\langle A_{0}B_{1}A_{1}B_{2}\ldots A_{R-1}B_{R}\right\rangle ,\\
\left\langle \sigma_{0}^{y}\sigma_{R}^{y}\right\rangle  & = & (-1)^{R-1}\left\langle B_{0}A_{1}B_{1}A_{2}\ldots B_{R-1}A_{R}\right\rangle ,\end{eqnarray*}
 where $A_{i}=c_{i}-c_{i}^{\dagger}$ and $B_{i}=c_{i}+c_{i}^{\dagger}$.
We can check that $\left\langle A_{0}A_{i}\right\rangle =\left\langle B_{0}B_{i}\right\rangle =0$
when $i\neq0$, and the complicated expression in the brackets above
can be handled with the help of Wick theorem \cite{barouch_statistical_1971,thermal_discord_2010}.
Let $G_{j-i}=\left\langle A_{i}B_{j}\right\rangle $, we have\begin{eqnarray*}
\left\langle \sigma^{z}\right\rangle  & = & G_{0},\\
\left\langle \sigma_{0}^{z}\sigma_{R}^{z}\right\rangle  & = & G_{0}^{2}-G_{R}G_{-R},\\
\left\langle \sigma_{0}^{x}\sigma_{R}^{x}\right\rangle  & = & \left|\begin{array}{cccc}
G_{-1} & G_{-2} & \ldots & G_{-R}\\
G_{0} & G_{-1} & \ldots & G_{-(R-1)}\\
\vdots & \vdots & \ddots & \vdots\\
G_{R-2} & G_{R-3} & \ldots & G_{-1}\end{array}\right|,\\
\left\langle \sigma_{0}^{y}\sigma_{R}^{y}\right\rangle  & = & \left|\begin{array}{cccc}
G_{1} & G_{0} & \ldots & G_{-(R-2)}\\
G_{2} & G_{1} & \ldots & G_{-(R-3)}\\
\vdots & \vdots & \ddots & \vdots\\
G_{R} & G_{R-1} & \ldots & G_{1}\end{array}\right|.\end{eqnarray*}
 And we have \begin{eqnarray*}
G_{R} & = & \left\langle A_{0}B_{R}\right\rangle =\left\langle (c_{0}-c_{0}^{\dagger})(c_{R}+c_{R}^{\dagger})\right\rangle \\
 & = & \frac{1}{N}\sum_{p,q=-N/2}^{N/2}e^{i\frac{2\pi q}{N}R}e^{i(\theta_{p}+\theta_{q})}\left\langle (\gamma_{p}^{\dagger}-\gamma_{-p})(\gamma_{-q}^{\dagger}+\gamma_{q})\right\rangle \\
 & = & -\frac{1}{N}\sum_{p}e^{i\frac{2\pi p}{N}R}e^{2i\theta_{p}}\\
 & \stackrel{N\rightarrow\infty}{=} & -\frac{1}{4\pi}\int_{-2\pi}^{2\pi}dr\frac{e^{\frac{i}{2}Rr}(B-e^{-ir})}{(1+B^{2}-2B\cos r)^{1/2}}.\end{eqnarray*}
Now we can get the expressions of correlation functions above back
into Eq. (\ref{eq:RDM}) and we have the reduced density matrix of
any two spins in the system. Bellow, we will discuss the pairwise
correlations in the system.

\subsection{Local correlations in quantum phase transition}

\begin{figure}
\includegraphics[width=7cm]{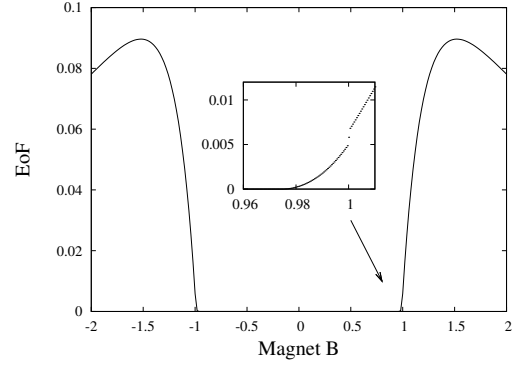}

\caption{The EoF is shown as a function of the magnetic field strength $B$.
The entanglement is born at $B_{E}\simeq0.9767$, and a sudden change
happens at the critical point. }

\label{Fig:Eof} 
\end{figure}

Now we discuss the correlations in the system. First, we calculate
the EoF of two local spins. As we know, the nearest two spins are
irrelevant. We give the EoF of the next-nearest spins shown in Fig.
\ref{Fig:Eof}. In fact, numerical results show that the EoF of the
two spins, whose distance is further than 2, is zero. 

In Sec. II, we noted that the quantum phase transition can be characterized
by SOP deduced from duality mapping, and the critical point lies at
$B=\pm1$. We can see that the EoF in {}``most'' of the topological
order area is zero and behaves like an order parameter, which is similar
to the logarithmic negativity in previous work \cite{kay_quantum_2004}.
However, the EoF is born before reaching $B=\pm1$, at the point around
$\left|B\right|\simeq0.9767$. There is a tiny {}``gap'' at the
critical point, which results from the finite scale, and that would
eliminate to a singular point in the thermodynamical limit.

\begin{figure*}
(a)%
\begin{minipage}[c]{2.4in}%
\includegraphics[bb=60bp 60bp 390bp 260bp,clip,width=6.5cm]{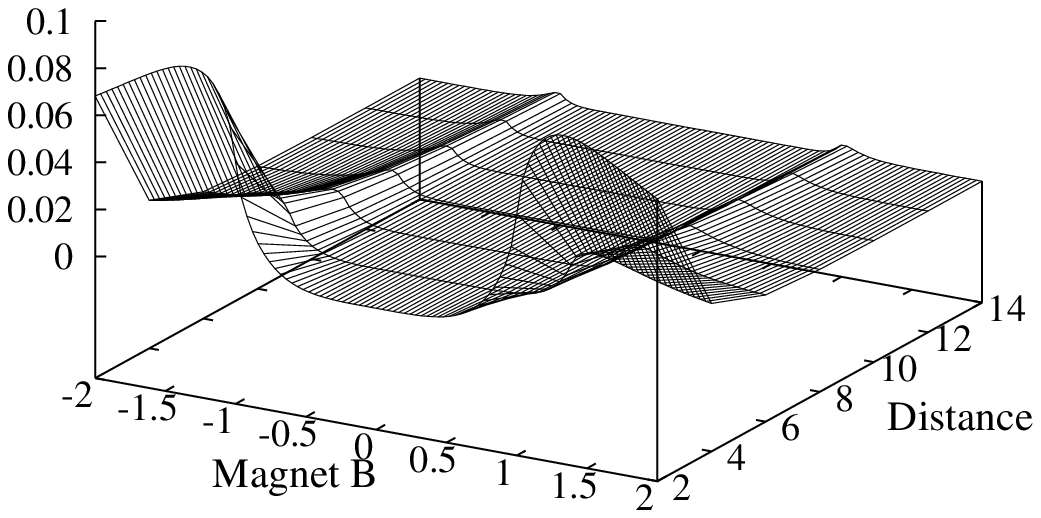}%
\end{minipage}\hspace{1.5cm}(b) %
\begin{minipage}[c]{2.4in}%
\includegraphics[width=6.5cm]{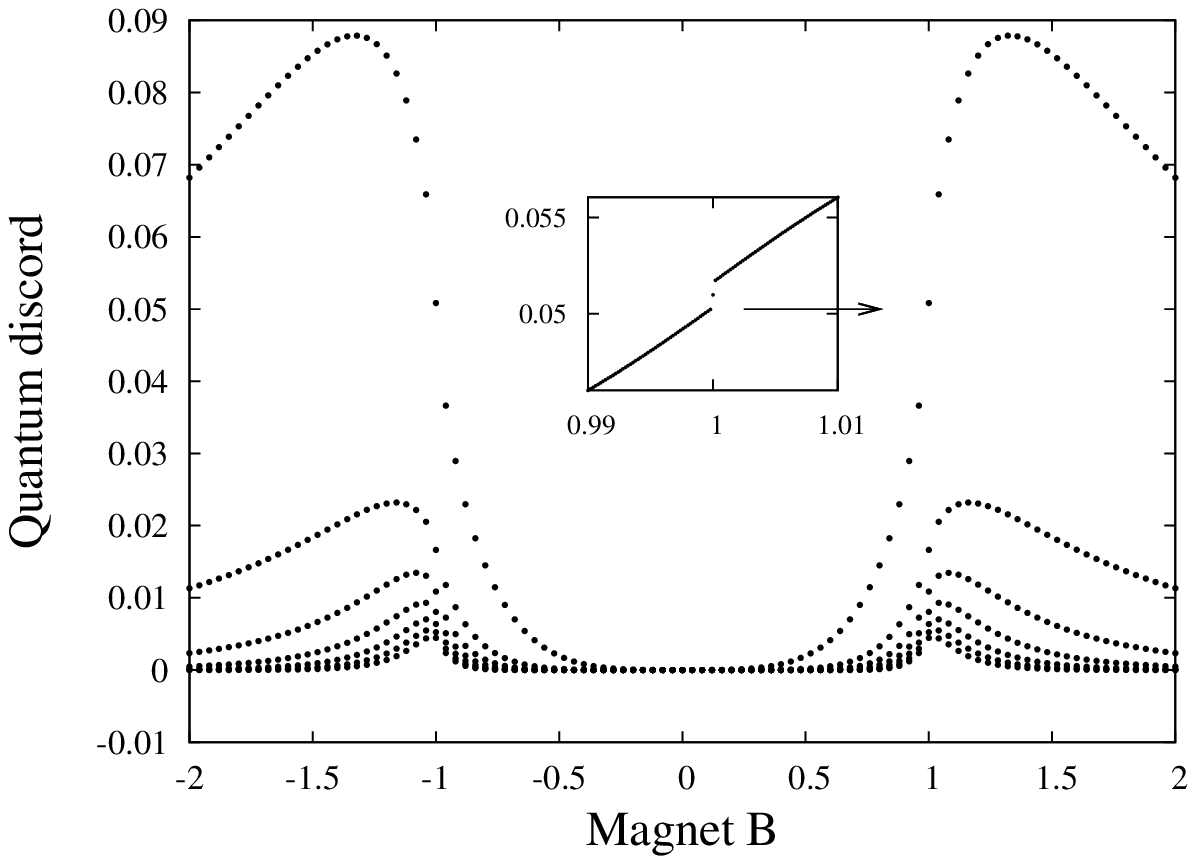}%
\end{minipage}

\caption{The quantum discord is plotted versus $B$ and $R$ where $B$ is
the magnetic field strength parameter and $R$ is the site's number
which correlates with site $0$.}

\label{Fig:discord} 
\end{figure*}

We cannot treat EoF as an order parameter. However, it tells us that
in the topological order area, local bipartite entanglement, as an
important quantum correlation, is greatly suppressed. It invokes us
to study the total quantum correlations in this area.

Second, we calculate the quantum discord of two spins with distance
$R$ in different magnetic field (Fig. \ref{Fig:discord}), where
$R$ is even. Around the point $B=1$, quantum discord has a tiny
gap similar to that of EoF. These behaviours both root in the property
of correlation functions and would eliminate to a singular point in
the thermodynamical limit. 

\begin{figure*}
(a)%
\begin{minipage}[c]{2.4in}%
\includegraphics[width=6cm]{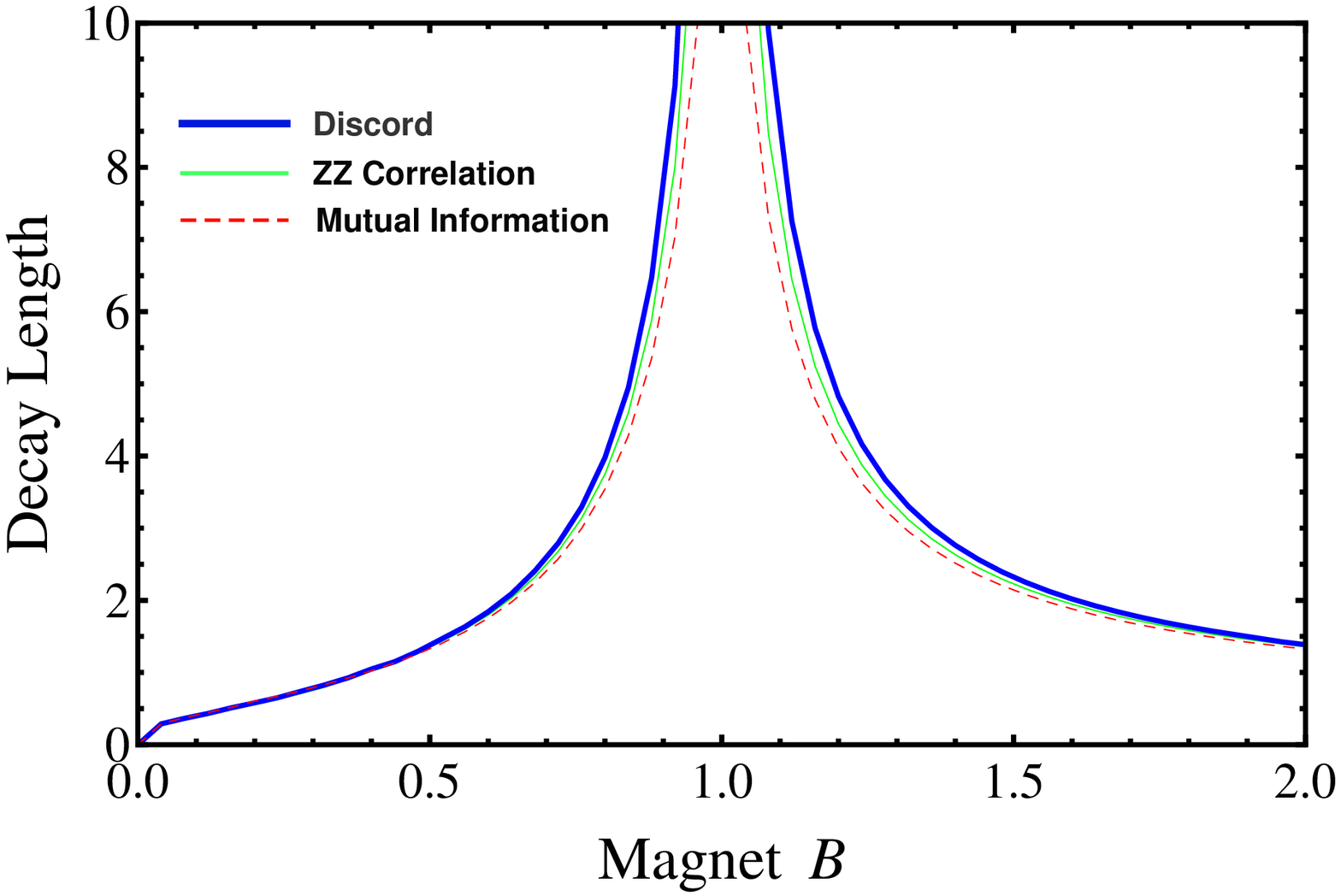}%
\end{minipage}\hspace{1.5cm}(b) %
\begin{minipage}[c]{2.4in}%
\includegraphics[width=6.4cm]{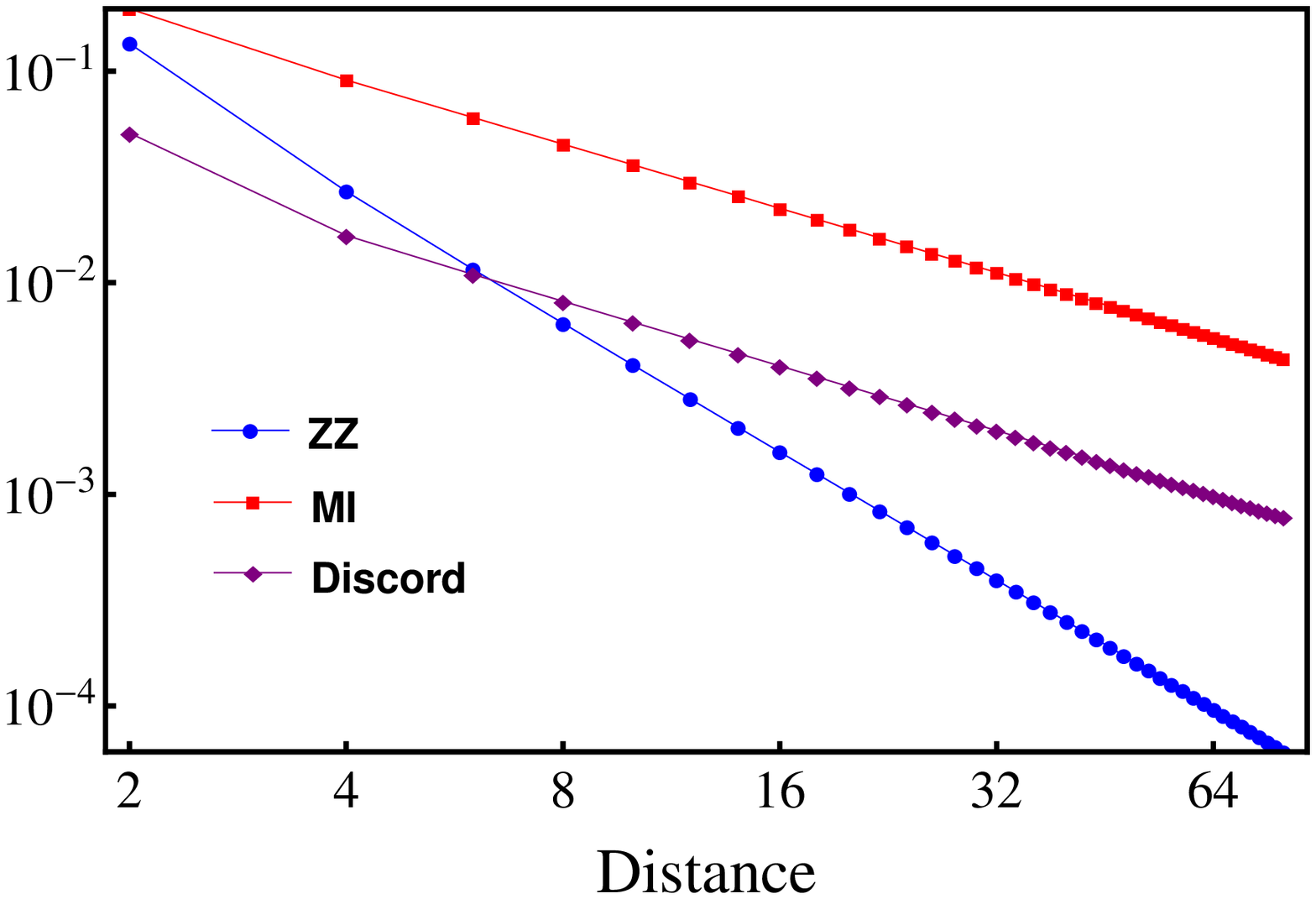}%
\end{minipage}\caption{ (Color online) (a) The decay length of quantum correlations (quantum
discord, mutual information, and the correlation function $\left\langle \sigma_{0}^{z}\sigma_{R}^{z}\right\rangle $)
are plotted versus magnetic field strength $B$. (b) The decay behaviour
at $B=1$. The correlations decay as $\sim R^{-\xi}$. We all take
$R$ as even.}

\label{Fig:decay} 
\end{figure*}

It was mentioned in Ref. \cite{chen_quantum_2010} that in a 2D TQPT,
local correlations are always classical and the quantum correlations
hides in the whole lattice, which also happens in many other 2D TQPT
systems. But things are different in 1D system (see Fig. \ref{Fig:discord}).

In 2D topological order systems, there often exist many different
conservative string operators whose paths are topologically equivalent,
and we can always find one that anti-commutes with certain local observable.
Therefore, most local correlation functions would be eliminated and
the density matrix Eq. (\ref{eq:RDM}) become diagonalized.

However, in 1D systems, degrees of freedom are restricted. There are
not so many topologically equivalent conservative string operators
as in 2D. The 1D systems do not pocess such high symmetry as in 2D
systems, and many local correlation functions survive. The quantum
discord, which measures the quantumness of pairwise correlations,
only gives zero at the cluster state when $B=0$. Nevertheless, qualitatively
speaking, we can see that the quantum discord is still quite small
in most of the topological order area compared with that in the area
$\left|B\right|>1$. And we say that the local quantum correlation
is greatly suppressed in the topological phase area.

On the other hand, this means in TQPT systems the global difference
of topology induced by dimension is reflected in the local quantum
correlations. The dimension constrains the topology of the system,
and also the types of global conservative quantities. In systems with
higher dimension like Ref. \cite{chen_quantum_2010}, the external
field term breaks some global conservative operators, while the survival
ones are still capable to eliminate local quantum correlations. However,
in 1D systems like what we study in this paper, there are not enough
global conservative quantities left in th presence the magnetic field
and the local quantum correlations are just suppressed. The survival
of local quantum correlation reflects the global restriction of the
topolgy induced by dimension.

Besides, we are interested in the decay behaviour of quantum discord
along with the increase of the distance of the two spins we study.
Numerical results show that the decay behaviours of the quantum discord
and total mutual information Eq. (\ref{eq:mi}) are just similar to
that of two-point correlation functions, i.e., they decay exponentially
when $\left|B\right|\neq1$ and with reversed power law at the critical
points. This is different from the sudden change behaviour of EoF,
although EoF and quantum discord are defined in a similar way, namely,
by finding the extremum. We show the exponential decay length of the
correlations with the magnetic field strength $B$ in Fig. \ref{Fig:decay}.

At the critical points $B=\pm1$, the correlations diverge as $\sim R^{-\xi}$.
We show them in Fig. \ref{Fig:decay}(b). For quantum discord, $\xi_{D}\simeq1.0576$
and mutual information $\xi_{M}\simeq1.0179$, and for the correlation
fucntion $\left\langle \sigma_{0}^{z}\sigma_{R}^{z}\right\rangle $,
$\xi_{ZZ}\simeq2.0464$. We guess this may relate to the universal
scaling factor. When $B=0$, the system is the cluster state. All
local quantum correlations vanish while quantum correlations still
hide in the chain globally.

\section{Summary}

We investigate a special model whose Hamiltonian contains three-spin
interactions. This model is composed of a cluster and a magnetic term,
and we discuss the topological properties of this system. The degeneracy
of the ground space differs in closed and open boundary conditions,
and the degeneracy is topologically protected. We obtained the global
SOP of this system by the method of duality mapping to characterize
the TQPT.

Further, we discuss quantum correlations of this system. We calculate
the quantum discord, mutual information and entanglement in this system.
The EoF of two local spins is {}``dead'' in most of the topological
order area. Together with the study of quantum discord, we believe
the quantum correlation is greatly suppressed in the topological order
area. This is different from previous work in 2D TQPT \cite{chen_quantum_2010},
where local quantum correlations all vanish. We believe that is because
in 1D systems, there is not so rich topology or high symmetry as in
2D systems. 

On the other hand, in topological order systems, the dimension of
the configuration constrains the topology of global conservative quantities.
This global difference of topology induced by dimension can be reflected
in the local quantum correlations. For example, the local quantum
correlations survive in 1D TQPT systems, while completely vanish in
2D, where there are more global conservative quantities left which
root in the richer topology of 2D systems.

Besides, we study the divergency behaviour of the correlations. The
quantum discord and mutual information diverge in reversed power law
at the critical points and exponentially elsewhere. We believe more
work can be done on the study of the universal scaling behaviour of
the divergency.
\begin{acknowledgments}
The work is supported in part by the NSF of China Grant No. 10775116,
No. 11075138, and 973-Program Grant No. 2005CB724508. 
\end{acknowledgments}

\end{document}